\documentclass[12pt]{article}

\begin{document}
\begin{center}

{\bf{\Large  Rotating Anisotropic Fluid Solutions
}}

\vspace{1cm}
E. Kyriakopoulos\footnote{E-mail: kyriakop@central.ntua.gr}\\
Department of Physics\\
National Technical University\\
157 80 Zografou, Athens, GREECE
\end{center}

\begin {abstract}
  An exact rotating anisotropic fluid solution and a family of exact rotating anisotropic fluid solutions are presented  which satisfy all energy conditions for certain values of their  parameters. The components of the Ricci tensor the eigenvalues of this tensor and the energy-momentum tensor of the solutions are given explicitly. All have the ring singularity of Kerr's solution and in addition the solution one more singularity and some solutions of the family additional singularities.The solution matches to the extremal solution of Kerr on two surfaces, which are thin shells and for proper values of the parameters of the solution approximate oblate spheroids. One of these  surfaces has positive surface density. The solutions of the family satisfy the matching conditions with the solution of Kerr on two pair of surfaces, which are again thin shells. The surface density of one pair of surfaces is given explicitly. Also for proper values of the parameters of the solutions the surfaces of the other pair approximate oblate spheroids.
\\

PACS number(s): 04.20.-q,  04.20.Jb

\end {abstract}

\section{Introduction}

Anisotropic fluid solutions play an important role in general theory of relativity \cite{Bo1}, \cite{He1}, even thought local isotropy is usually assumed. Such solutions may be useful in connection with the important problem of finding internal solutions of Kerr's solution \cite{Ke1}. Most efforts in this direction were based on models with regular disks most of which \cite{Ha1} but not all \cite{Bi1} seem to have serious problems. The attempts to find internal solutions of perfect fluid type were not successful \cite{Kr1}, \cite{Mc1}. Therefore it is natural to turn to anisotropic fluid solutions. In a previous paper we found a family of anisotropic fluid solutions which match to Kerr's solution \cite{Ky1}. In this paper we present an exact anisotropic fluid solution and a family of exact anisotropic fluid solutions and discuss their matching to Kerr's solution.

In Sect. 2 we start from a metric $g_{\mu\nu}$ whose form approaches the general form of a metric we have considered before \cite{Ky2} and has two functions $h(r)$ and $f(r,x)$ to be determined. From this metric the non-zero components of the Ricci tensor $R_{\mu\nu}$ are determined and then the expression, from which the eigenvalues of the matrix $R_{\mu}^{\nu}$ can be found, is given. This expression is written in two forms. Then we find equations the satisfaction of which makes one of the roots $\lambda_t$ and $\lambda_\phi$ of $R_\mu^\nu$ equal to one of the two other roots $\lambda_r$ and $\lambda_\theta$ of $R_\mu^\nu$. We distinguish six cases in which this happens and give the equations which must be satisfied in each case. Then we find solutions of the equations in
each case. A solution is anisotropic if the eigenvalues of $R_\mu^\nu$ satisfy the condition \cite{He1} \cite{Le1}
\begin{equation}
\lambda_\theta=\lambda_\phi  \label{1-1}
\end{equation}

Solving equations of case 1 we find the family of anisotropic fluid solutions which we presented in Ref \cite{Ky1} and in addition one more solution which satisfies Eq (\ref{1-1}) and is therefore an anisotropic fluid solution. The solutions we get by solving equations of case 3 are special cases of the solutions of case 1.

Solving equations of case 4 we find a family of solutions which are not anisotropic fluid solutions, while the solutions of equations of case 2 are special cases of the solutions of equations of case 4. Also the solution we get by solving equations of case 5 is not anisotropic. Finally solving equations of case 6 we find a family of solutions which satisfy Eq (\ref{1-1}) and are therefore anisotropic.

In Sect III the non zero components of the Ricci tensor $R_{\mu\nu}$ of the anisotropic fluid solution we found in Sect II are given explicitly and the eigenvalues $\lambda_i$,  $ i=t, r, \theta, \phi$ of the matrix $R_\mu^\nu$ are obtained. It is shown that the solution satisfies the dominant the strong and therefore the weak energy  energy conditions \cite{Ha2} for some values of its parameters. The eigenvectors, which correspond to the eigenvalues of $R_\mu^\nu$, are calculated and the energy-momentum tensor $T_{\mu\nu}$ is given explicitly. As expected this tensor has the standard form of the energy-momentum tensor of an anisotropic fluid solution\cite{He1} \cite{Le1}. It is found that this anisotropic fluid solution has the well known ring singularity of Kerr's solution \cite{Ra1} and in addition one more singularity. The infinite red shift surfaces of the solution are given.

The matching of the anisotropic fluid solution to Kerr's solution is discussed using the Darmois-Israel boundary conditions \cite{Do1}, \cite{Is1}. From the continuity of the first fundamental form we get two matching surfaces, which for small value of a constant approximate oblate spheroids. Also it is found that all components of the extrinsic curvature except one are continuous across the surfaces, which  means that the surfaces are thin shells. The non zero components of the surface energy tensor are found and from these the surface density is calculated. It is found that for one of the matching surfaces the surface density is positive everywhere. We should point out that the exterior solution is the extremal Kerr's solution and that the solutions of the family of Ref \cite{Ky1} do not match to the extremal solution of Kerr.

In Sect IV the non zero components of the Ricci tensor $R_{\mu\nu}$ of the family of solutions of case 6 are given and the eigenvalues $\lambda_i$,  $ i=t, r, \theta, \phi$ of the matrix $R_\mu^\nu$ are calculated and found to satisfy Eq (\ref{1-1}). Also the eigenvectors which corresponds to the eigenvalues $\lambda_i$ are calculated and the energy-momentum tensor $T_{\mu\nu}$ is given explicitly. The solutions of the family satisfy
the dominant the strong and therefore the weak energy conditions \cite{Ha2}, if a relation is satisfied. It is found that all solutions of the family have the ring singularity of Kerr's solution \cite{Ra1} and some of them additional singularities.Also it is shown that the solutions of the family satisfy the matching conditions with the solution of Kerr on two pair of surfaces. Since not all components of the extrinsic curvature are continuous on these surfaces all surfaces are thin shells. For two of them we calculated the discontinuity of the extrinsic curvature on them, the surface energy tensor and also the surface density. The other two surfaces for proper values of the parameters of the solution approximate oblate spheroids and one of them is inside the outer red shift surface and inside the outer event horizon of the exterior solution of Kerr or coincides with them.

\section{Model's equations and their solutions}

Consider a metric which in Boyer-Lindquist coordinates \cite{Bo2} has the form
\[g_{\mu\nu}=\{[-T, 0, 0, -a(1-x^2)(1-T)], [0, \frac{f(r,x)}{\rho^2T+a^2(1-x^2)}, 0, 0],\]
\begin{equation}
[0, 0, f(r,x), 0], [-a(1-x^2)(1-T), 0, 0, (1-x^2)(r^2+a^2x^2+a^2(1-x^2)(2-T))]\}
\label{2-1}
\end{equation}
where a is an arbitrary constant,
\begin{equation}
x=\cos\theta,\>\>\>\>\>\>\>\> \rho^2=r^2+a^2x^2, \>\>\>\>\>\>\>\>T=1+\frac{h(r)}{\rho^2}
\label{2-2}
\end{equation}
and h(r) and f(r,x) are functions to be determined. The above $g_{\mu\nu}$ is the same with the $g_{\mu\nu}$ of Ref
\cite{Ky1}, approaches the $g_{\mu\nu}$ of Ref \cite{Ky2} and becomes the metric of Kerr \cite{Ke1} if
\begin{equation}
 h(r)=-2Mr \>\>\>\mbox{and}\>\>\> f(r,x)=\rho^2
\label{2-3}
\end{equation}
To make the calculations easier we shall replace $f(r,x)$ with a new function $s(r,x)$ by the relation
\begin{equation}
 f(r,x)=b \rho^2 e^{-s(r,x)}
\label{2-4}
\end{equation}
where $b$ is a constant. From the above $g_{\mu\nu}$ we can calculate the Ricci tensor $R_{\mu\nu}$  \cite{Bo3} and then the eigenvalues $\lambda_i$,  $ i=t, r, \theta, \phi$ of the matrix $R_\mu^\nu$. To proceed let us denote by prime the differentiation of $h(r)$ with respect to $r$ and by a lower index $r$ or $x$ the differentiation of $s(r,x)$ with respect to $r$ or $x$,
 for example
\begin{equation}
 h''(r)=\frac{d^2h(r)}{dr^2} \>\>\>\mbox{and}\>\>\> {s_r}(r,x)=\frac{\partial s(r,x)}{\partial r}
\label{2-5}
\end{equation}
Also let us introduce the notation
\begin{equation}
 \lambda_1=\frac{ e^{-s(r,x)}}{ b (\rho^2)^2}\{h(r)- r h'(r)\}
\label{2-7}
\end{equation}
\begin{equation}
 \lambda_2=-\frac{ e^{-s(r,x)}}{2 b (\rho^2)^2}\{2h(r)-2 r h'(r)+\rho^2h''(r)\}
\label{2-8}
\end{equation}
\begin{equation}
\Gamma_\pm(r,x)=(1-x^2)(2 r+h'(r))s_x\pm2 x(r^2+a^2+h(r))s_r(r,x)
\label{2-9}
\end{equation}
\begin{equation}
E(r,x)=(1-x^2)s_{xx}(r,x) + (r^2+a^2+h(r))s_{rr}(r,x)
\label{2-10}
\end{equation}
Then we find that the eigenvalues $\lambda_i$ are the solutions of equation
\[
Det(R_\mu^\nu-\lambda\delta_\mu^\nu)=(\lambda-\lambda_1)(\lambda-\lambda_2)\{-\frac{ e^{-2 s(r,x)}}{16 b^2 (1-x^2)(\rho^2)^2(r^2+a^2+h(r))}\Gamma_-^2+\]
\[\{\lambda-\lambda_1+\frac{ e^{-s(r,x)}}{2 b \rho^2}[E(r,x)+(2 r+h'(r))s_r(r,x)]\}\]
\begin{equation}
\{\lambda-\lambda_2+\frac{ e^{-s(r,x)}}{2 b \rho^2}[E(r,x)-2x s_x(r,x)]\}\}=0
\label{2-11}
\end{equation}
and the root $\lambda_1$ is one of the roots $\lambda_t$ and $\lambda_\phi$ and the root $\lambda_2$ the other.
Also the expression $Det(R_\mu^\nu-\lambda\delta_\mu^\nu)$ can be written in the form
\[
Det(R_\mu^\nu-\lambda\delta_\mu^\nu)=(\lambda-\lambda_1)(\lambda-\lambda_2)\{-\frac{ e^{-2 s(r,x)}}{16 b^2 (1-x^2)(\rho^2)^2(r^2+a^2+h(r))}\Gamma_+^2+\]
\[\frac{ e^{-s(r,x)}}{2 b \rho^2}E(r,x)\{\frac{ e^{-s(r,x)}}{2 b \rho^2}E(r,x)+2\lambda-\lambda_1-\lambda_2+\]\[\frac{ e^{-s(r,x)}}{2 b \rho^2}[(2 r+h'(r))s_r(r,x)-2 x s_x(r,x)]\}+\]
\begin{equation}
(\lambda-\lambda_1)(\lambda-\lambda_2)+\frac{ e^{-s(r,x)}}{2 b \rho^2}[(\lambda-\lambda_2)(2 r+h'(r))s_r(r,x)-(\lambda-\lambda_1)2 x s_x(r,x)]\}
\label{2-12}
\end{equation}

A solution is anisotropic if Eq  (\ref{1-1}) holds. Eqs (\ref{2-11}) and  (\ref{2-12}) have the roots $\lambda_1$
and $\lambda_2$ and two more, which let us call $\lambda_3$ and $\lambda_4$. We shall try to find solutions for which  Eq (\ref{1-1}) holds. To do that we shall try first to find solutions for which one of the roots $\lambda_1$ or $\lambda_2$ is equal to one of the roots $\lambda_3$ or $\lambda_4$ and then we shall check if Eq (\ref{1-1}) is satisfied. Consider first Eq (\ref{2-11}). The root $\lambda_1$ of this equation is equal to one of the roots $\lambda_3$ or $\lambda_4$ if the following relations hold:

Case 1
\begin{equation}
\Gamma_-(r,x)=0\>\>\>\mbox{and}\>\>\>E(r,x)+(2 r+h'(r))s_r(r,x)=0
\label{2-13}
\end{equation}

Case 2

\[\Gamma_-(r,x)=0\>\>\>\mbox{and}\>\>\>\rho^2(E(r,x) -2 x s_x(r,x))+\]
\begin{equation}
4(h(r)-r h'(r))+\rho^2h''(r)=0
\label{2-14}
\end{equation}
while the root $\lambda_2$ is equal to one of the roots $\lambda_3$ or $\lambda_4$ in the following two cases:

Case 3

\[\Gamma_-(r,x)=0\>\>\>\mbox{and}\>\>\>\rho^2[E(r,x)+(2 r+h'(r))s_r(r,x)] \]
\begin{equation}
-4(h(r)-r h'(r))-\rho^2h''(r)=0
\label{2-15}
\end{equation}

Case 4

\begin{equation}
\Gamma_-(r,x)=0\>\>\>\mbox{and}\>\>\>E(r,x)-2 x s_x(r,x)=0
\label{2-16}
\end{equation}
Consider now Eq (\ref{2-12}). Its root $\lambda_1$  is equal to one of the roots $\lambda_3$ or $\lambda_4$ if
\begin{equation}
\Gamma_+(r,x)=0
\label{2-16'}
\end{equation}
and
\[E(r,x)\{\rho^2[E(r,x)+(2 r+h'(r))s_r(r,x)-2xs_x(r,x)]+4(h(r)-r h'(r))+\]
\begin{equation}
\rho^2h''(r)\}+(2r+h'(r))[4(h(r)-r h'(r))+\rho^2h''(r)]s_r(r,x)=0
\label{2-16''}
\end{equation}
We shall consider solutions for which
\begin{equation}
(2r+h'(r))[4(h(r)-r h'(r))+\rho^2h''(r)]=0
\label{2-16'''}
\end{equation}
 This relation is satisfied if $h(r)=-r^2+w$ and if $h(r)=-2Mr$ where w and M are arbitrary constants. But if $h(r)=-r^2+w$ Eq (\ref{2-16'}) gives $w=a^2$, which makes the metric singular since for $h(r)=-r^2+a^2$ we have $\rho^2T+a^2(1-x^2)=0$. This means that we must take $h(r)=-2Mr$, in which case we get $\lambda_1=\lambda_2=0$ and Eq (\ref{2-16''}) becomes
\begin{equation}
E(r,x)\{E(r,x)+2(r-M)s_r(r,x)-2xs_x(r,x)\}=0
\label{2-17}
\end{equation}
Therefore the roots $\lambda_1$, $\lambda_2$ and one of the roots $\lambda_3$ and $\lambda_4$ vanish in the following two cases:

Case 5

\begin{equation}
\Gamma_+(r,x)=0\>\>\>\mbox{and}\>\>\>E(r,x)=0\>\>\>\mbox{and}\>\>\>h(r) =-2 M r
\label{2-17}
\end{equation}

 Case 6

\[\Gamma_+(r,x)=0\>\>\>\mbox{and}\>\>\>E(r,x)+2(r-M)s_r(r,x)-2 x s_x(r,x)=0\]
\begin{equation}
\>\>\>\mbox{and}\>\>\>h(r) =-2 M r
\label{2-18}
\end{equation}
where $M$ is an arbitrary constant. We shall find solutions for all cases.

The first of Eqs (\ref{2-13}) is satisfied if $h(r)$ is an arbitrary function of $r$ and

\begin{equation}
s(r,x)=F[\frac{1-x^2}{r^2+a^2+h(r)}]
\label{2-19}
\end{equation}
where $F$ is an arbitrary function of its arguments. Substituting the above expression in the second of Eqs (\ref{2-13}) and solving the resulting relation we find two solutions for $s(r,x)$ and $h(r)$. The solutions expressed in terms of $f(r,x)$ of Eq (\ref{2-4}) and $h(r)$ are the following:
\begin{equation}
f(r,x)=b \rho^2[\frac{(r-M)^2+kx^2}{(r-M)^2+k}]^c
\label{2-20}
\end{equation}
\begin{equation}
h(r)=-2 M r+M^2-a^2+k
\label{2-21}
\end{equation}
where $a$, $b$, $c$, $M$, and $k$ are arbitrary constants and $k\neq0$ and
\begin{equation}
f(r,x)=b \rho^2 e^{c\frac{x^2-1}{(r-M)^2}}
\label{2-22}
\end{equation}
\begin{equation}
h(r)=-2 M r+M^2-a^2
\label{2-23}
\end{equation}
where  $a$, $b$, $c$ and $M$ are arbitrary constants.The metric of Eqs (\ref{2-1}), (\ref{2-20}) and (\ref{2-21}) is the metric of the family of anisotropic fluid solutions of Ref. \cite{Ky1}. Also we
 shall show in Sect III that  the solution given by Eqs (\ref{2-1}), (\ref{2-22}) and (\ref{2-23}) is a rotating anisotropic fluid solution.

 Solutions of  Eqs (\ref{2-15}) of case 3 are the family of solutions of Eqs (\ref{2-20}) and (\ref{2-21}) of case 1 with $k=a^2-M^2$ and the solution of  Eqs (\ref{2-22}) and (\ref{2-23}) of case 1 with  $a^2-M^2=0$.
Therefore the solutions of case 1 include the solutions of case 3.

Solving Eqs (\ref{2-16}) of case 4 and expressing the solution in terms of $f(r,x)$ of Eq (\ref{2-4}) we get
\begin{equation}
f(r,x)=b\rho^2[\frac{1-x^2}{(r-M)^2+k x^2}]^c
\label{2-24}
\end{equation}
\begin{equation}
h(r)=-2 M r+M^2-a^2+k
\label{2-25}
\end{equation}
where $a$, $b$, $c$, $M$, and $k$ are arbitrary constants. The eigenvalues $\lambda_i$ $i=t, r, \theta, \phi$ of the tensor $R_\mu^\nu$, which comes from this metric, are the following:
\begin{equation}
\lambda_t=\lambda_r=-\lambda_\phi=\frac{a^2-M^2-k}{b \rho^2}[\frac{1-x^2}{(r-M)^2+k x^2}]^{-c}
\label{2-26}
\end{equation}
\begin{equation}
\lambda_\theta=\frac{2c \rho^2-(a^2-M^2-k)(1-x^2)}{b(1-x^2)\rho^2}[\frac{1-x^2}{(r-M)^2+k x^2}]^{-c}
\label{2-27}
\end{equation}
Therefore Eq (\ref{1-1}) is not satisfied and the solutions of this family are not anisotropic. The expressions for $f(r,x)$ and $h(r)$ we find by solving Eqs (\ref{2-14}) of case 2 are those of Eqs (\ref{2-26}) and (\ref{2-27}) with $k=a^2-M^2$. Therefore the family of solutions of case 2 is obtained from the family of solutions of case 4 for $k=a^2-M^2$.

The general solution of the first of Eqs (\ref{2-18}) of case 6, with $h(r)=-2 M r$ according to the third of Eqs
(\ref{2-18}), is
\begin{equation}
s(r,x)=H[(1-x^2)(r^2-2 M r+a^2)]
\label{2-28}
\end{equation}
where $H$ an arbitrary function of its arguments. Substituting the above expression for $s(r.x)$ in the second of Eqs (\ref{2-18}) and solving the resulting relation, having in mind that $h(r)=-2 M r$, we can calculate $s(r,x)$. Expressing the solution in terms of $f(r,x)$ of Eq (\ref{2-4}) we find for case 6
\begin{equation}
f(r,x)=b\rho^2[(1-x^2)(r^2-2 M r+a^2)]^c \>\>\>\mbox{and}\>\>\>h(r)=-2 M r
\label{2-29}
\end{equation}
where  $a$, $b$, $c$ and $M$ are arbitrary constants. In Sect IV we shall show that the family of solutions with metric given by Eqs (\ref{2-1}) and (\ref{2-29}) is a family of anisotropic fluid solutions.

Finally solving Eqs (\ref{2-17}) and using Eq (\ref{2-4}) we find for case 5 the solution
\begin{equation}
f(r,x)=b\rho^2 e^{c(1-x^2)(r^2-2 M r+a^2)} \>\>\>\mbox{and}\>\>\>h(r)=-2 M r
\label{2-30}
\end{equation}
where  $a$, $b$, $c$ and $M$ are arbitrary constants. The eigenvalues $\lambda_i$   $i=t, r, \theta, \phi$ of the tensor $R_\mu^\nu$ of this solution are
\begin{equation}
\lambda_t=\lambda_r=\lambda_\phi=0
\label{2-31}
\end{equation}
\begin{equation}
\lambda_\theta=-\frac{2 c}{b \rho^2}[(r-M)^2+(a^2-M^2)x^2)]e^{-c(1-x^2)(r^2-2 M r+a^2)}
\label{2-32}
\end{equation}
Therefore this is not an anisotropic fluid solution.

\section{Rotating anisotropic fluid solution }

The non zero components of the Ricci tensor $R_{\mu\nu}$ of the metric $g_{\mu\nu}$ of Eq (\ref{2-1}) with $f(r,x)$ and $h(r)$ given by Eqs (\ref{2-22}) and (\ref{2-23}) are the following \cite{Bo3}
\begin{equation}
R_{tt}=\frac{(M^2-a^2)[(r-M)^2+a^2(1-x^2)]}{b(\rho^2)^3}e^L
\label{3-1}
\end{equation}
\begin{equation}
R_{rr}=\frac{-(M^2-a^2)(r-M)^2+2 c \rho^2}{(r-M)^4\rho^2}
\label{3-2}
\end{equation}
\begin{equation}
R_{\theta\theta}=\frac{M^2-a^2}{\rho^2}
\end{equation}
\begin{equation}
R_{\phi\phi}=\frac{(M^2-a^2)(1-x^2)[(r^2+a^2)^2+a^2(r-M)^2(1-x^2)]}{b(\rho^2)^3}e^L
\label{3-4}
\end{equation}
\begin{equation}
R_{t\phi}=-\frac{a(M^2-a^2)(1-x^2)[(r-M)^2+r^2+a^2]}{b(\rho^2)^3}e^L
\label{3-5}
\end{equation}
where
\begin{equation}
L=c\frac{1-x^2}{(r-M)^2}
\label{3-6}
\end{equation}
Also the Ricci scalar of the solution is
\begin{equation}
R=\frac{2 c}{b(r-M)^2\rho^2}e^L
\label{3-7}
\end{equation}
From Eqs (\ref{3-1}) - (\ref{3-5}) and the metric $g_{\mu\nu}$ of the solution we can calculate the matrix $R_\mu^\nu$ and then its four eigenvalues, which are found to be the following:
\begin{equation}
\lambda_\pm=\pm\frac{M^2-a^2}{b(\rho^2)^2}e^L
\label{3-8}
\end{equation}
\begin{equation}
\lambda_r=\frac{-(M^2-a^2)(r-M)^2+2c\rho^2}{b(r-M)^2(\rho^2)^2}e^L
\label{3-9}
\end{equation}
\begin{equation}
\lambda_\theta=\frac{M^2-a^2}{b(\rho^2)^2}e^L=\lambda_+
\label{3-10}
\end{equation}
Therefore the eigenvalue $\lambda_t$ is one of the eigenvalues $\lambda_+$ or $\lambda_-$ and the eigenvalue $\lambda_\phi$ the other. To find which one is $\lambda_t$ and which $\lambda_\phi$ we calculate the eigenvectors which correspond to these eigenvalues. We find that the normalized timelike eigenvector $(u_t)^\mu$, which has the eigenvalue $\lambda_t$, is given by the relation
\begin{equation}
(u_t)^\mu=\frac{1}{(r-M)\sqrt{\rho^2}}[(r^2+a^2)\delta_t^\mu+a\delta_\phi^\mu]
\label{3-11}
\end{equation}
and corresponds to the eigenvalue $\lambda_-$. This means that
\begin{equation}
\lambda_t=\lambda_-=-\frac{M^2-a^2}{b(\rho^2)^2}e^L
\label{3-12}
\end{equation}
Therefore we must have
\begin{equation}
\lambda_\phi=\lambda_+=\lambda_\theta
\label{3-13}
\end{equation}
which means that our solution is a rotating anisotropic fluid solution. In expression (\ref{3-11}) and in all expressions for the eigenvectors which follow we have chosen the sign + as overall sign. Also we find that the normalized eigenvector $(u_\phi)^\mu$, which corresponds to the eigenvalue $\lambda_\phi=\lambda_+$, is given by the relation
\begin{equation}
(u_\phi)^\mu=\frac{1}{\sqrt{\rho^2(1-x^2)}}[a(1-x^2)\delta_t^\mu+\delta_\phi^\mu]
\label{3-14}
\end{equation}
Finally the normalized eigenvectors $(u_r)^\mu$ and $(u_\theta)^\mu$, which correspond to the eigenvalues $\lambda_r$ and $\lambda_\theta$ respectively are given by the relations
\begin{equation}
(u_r)^\mu=\frac{r-M}{\sqrt{b\rho^2}}e^{L/2}\delta_r^\mu
\label{3-15}
\end{equation}
\begin{equation}
(u_\theta)^\mu=\frac{1}{\sqrt{b\rho^2}}e^{L/2}\delta_\theta^\mu
\label{3-16}
\end{equation}
From  Eqs (\ref{3-11}) and  (\ref{3-14})-(\ref{3-16}) we find the normalized eigenvectors $(u_t)_\mu$, $(u_\phi)_\mu$, $(u_r)_\mu$ and $(u_\theta)_\mu$. We get
\begin{equation}
(u_t)_\mu=\frac{r-M}{\sqrt{\rho^2}}[-\delta_{t\mu}+a(1-x^2)\delta_{\phi\mu}]
\label{3-17}
\end{equation}
\begin{equation}
(u_\phi)_\mu=\sqrt{\frac{1-x^2}{\rho^2}}[-a\delta_{t\mu}+(r^2+a^2)\delta_{\phi\mu}]
\label{3-18}
\end{equation}
\begin{equation}
(u_r)_\mu=\frac{\sqrt{b\rho^2}}{r-M}e^{-L/2}\delta_{r\mu}
\label{3-19}
\end{equation}
\begin{equation}
(u_\theta)_\mu=\sqrt{b\rho^2}e^{-L/2}\delta_{\theta\mu}
\label{3-20}
\end{equation}

The eigenvalues $w_i=\lambda_i-R/2$, $i=t, r, \theta$ and $ \phi$ of the energy-momentum tensor $T_\mu^\nu=R_\mu^\nu-\frac{R}{2}\delta_\mu^\nu$ are calculated from Eqs (\ref{3-7}, (\ref{3-9}), (\ref{3-10}), (\ref{3-12}) and (\ref{3-13}). We get
\begin{equation}
w_t=-\frac{(M^2-a^2)(r-M)^2+c\rho^2}{b(r-M)^2(\rho^2)^2}e^L
\label{3-21}
\end{equation}
\begin{equation}
w_r=\frac{-(M^2-a^2)(r-M)^2+c\rho^2}{b(r-M)^2(\rho^2)^2}e^L
\label{3-22}
\end{equation}
\begin{equation}
w_\theta=w_\phi=\frac{(M^2-a^2)(r-M)^2-c\rho^2}{b(r-M)^2(\rho^2)^2}e^L
\label{3-23}
\end{equation}
Therefore the energy density $\mu$ is given by the relation
\begin{equation}
\mu=-w_t=\frac{(M^2-a^2)(r-M)^2+c\rho^2}{b(r-M)^2(\rho^2)^2}e^L
\label{3-24}
\end{equation}
If we introduce the notation
\begin{equation}
Z=\frac{(M^2-a^2)(r-M)^2}{b(r-M)^2(\rho^2)^2}e^L\>\>\>\mbox{and}\>\>\>Q=\frac{c}{b(r-M)^2\rho^2}e^L
\label{3-25}
\end{equation}
Eqs (\ref{3-22}) - (\ref{3-24}) become
\begin{equation}
\mu=Z+Q, \>\>\>w_r=-Z+Q,\>\>\>w_\theta=w_\phi=Z-Q
\label{3-26}
\end{equation}
It is easy to show that if
\begin{equation}
Z\geq0\>\>\>\mbox{and}\>\>\>Q\geq0
\label{3-27}
\end{equation}
the above expressions for $\mu $, $w_r$, $w_\theta$ and $w_\phi$ satisfy the dominant the strong and therefore the weak energy conditions \cite{Ha2}. Relations (\ref{3-27}) are satisfied if
\begin{equation}
\frac{M^2-a^2}{b}\geq0\>\>\>\mbox{and}\>\>\>\frac{c}{b}\geq0
\label{3-28}
\end{equation}
From  Eqs (\ref{3-1}) - (\ref{3-5}), (\ref{3-7}) and the metric $g_{\mu\nu}$ we can calculate the energy-momentum tensor $T_{\mu\nu}$. Writing
\begin{equation}
w_r=w_\parallel\>\>\>\mbox{and}\>\>\>w_\theta=w_\phi=w_\perp
\label{3-29}
\end{equation}
we find using Eqs (\ref{3-17}), (\ref{3-19}) and (\ref{3-22}) - (\ref{3-24}) that this tensor can be written in the form
\begin{equation}
T_{\mu\nu}=(\mu+w_\perp)(u_t)_\mu(u_t)_\nu+w_\perp g_{\mu\nu}+(w_\parallel-w_\perp)(u_r)_\mu(u_r)_\nu
\label{3-30}
\end{equation}
which is the form of the energy-momentum tensor of an anisotropic fluid  \cite{He1}, \cite{Le1}.

The infinite red shift surfaces of a solution are obtained from the relation $g_{tt}$=0. From Eqs (\ref{2-1}), (\ref{2-2}) and (\ref{2-23}) we get
\begin{equation}
r_\pm^{RS}=M \pm a\sqrt{1-x^2}
\label{3-31}
\end{equation}
The above surfaces are closed and axially symmetric.

A solution has irremovable singularities at the points at which at least one of the invariants $R$ and $R^2=R_{\mu\nu\zeta\eta}R^{\mu\nu\zeta\eta}$ is singular. The Ricci scalar $R$ is given by Eq (\ref{3-7}), while the curvature scalar $R^2$ is given by the relation \cite{Bo3}
\begin{equation}
R^2=\frac{U}{b^2(r-M)^4(\rho^6)^2}e^{2L}
\label{3-32}
\end{equation}
where $U$ is a complicated polynomial of $a$, $c$, $M$, $r$ and $x$. Therefore the solution has an irremovable singularity when
\begin{equation}
\rho^2=r^2+a^2x=0
\label{3-33}
\end{equation}
which is the well known ring singularity of Kerr's solution \cite{Ra1}, \cite{Vi1}, and also when
\begin{equation}
r=M
\label{3-34}
\end{equation}
The singularity at $r=M$ is for $x^2=1$ on the infinite red shift surfaces and for $x^2\neq1$ insides the outer red shift surface.

To examine if a solution can be matched to the solution of Kerr, which is expressed in Boyer-Lindquist coordinates \cite{Bo2}, on a surface $S$ we assume that the coordinates $r$ and $\theta$ of the various points of the surface are not independent but can be expressed with the help of a single parameter $\tau$. This means that the surface has coordinates $\zeta^i=(t,\tau,\phi)$ while the space-time has coordinates $x^\alpha=(t,r,\theta,\varphi)$. Eliminating $\tau$ we get for $S$ an equation of the form
\begin{equation}
r=R(\theta)
\label{3-35}
\end{equation}
One can show that the relation which connects the 3-metric $^3g_{ij}$ of the surface with the 4-metric $^4g_{\mu\nu}$ of the space-time is the following
\begin{equation}
^3g_{ij}=\frac{\partial x^\alpha}{\partial \zeta^i}\frac{\partial x^\beta}{\partial \zeta^j}{^4g_{\alpha\beta}}
\label{3-36}
\end{equation}
where Greek indices take the four values $(t,r,\theta,\phi)$ and Latin indices the three values $(t,\tau,\phi)$. If $P$ is a metric dependent quantity the notation $P^+ (P^-)$ means that $P$ is calculated in the exterior (interior) region of $S$, while the notation $P^+|_S $
 $ (P^-|_S)$ means that $P$ is calculated in the exterior (interior) region of $S$ and evaluated at the surface. The solution of Kerr occupies the exterior region. We use the notation
\begin{equation}
[P]\equiv P^{+}|_S-P^{-}|_S
\label{3-37}
\end{equation}
which means that [P] denotes the discontinuity of $P$ at the surface.

The Darmois-Israel boundary conditions \cite{Do1}, \cite{Is1} for the matching of the two regions exterior and interior are continuity the first fundamental form
\begin{equation}
[^3g_{ij}]=0
\label{3-38}
\end{equation}
and continuity of the extrinsic curvature $K_{ij}$ ( second fundamental form)
\begin{equation}
[K_{ij}]=0
\label{3-39}
\end{equation}
If both conditions are satisfied we refer to $S$ as boundary surface. If only condition (\ref{3-38}) is satisfied we refer to $S$ as thin shell. If $g_{t\phi}$ is the only non-zero off diagonal element of $g_{\mu\nu}$ Eq (\ref{3-38}) implies the relations \cite{Dr1}
\begin{equation}
[^3g_{tt}]=[^4g_{tt}]=0, \>\>\>\> [^3g_{t\phi}]=[^4g_{t\phi}]=0, \>\>\>\> [^3g_{\phi\phi}]=[^4g_{\phi\phi}]=0
\label{3-40}
\end{equation}
\begin{equation}
[^3g_{\tau\tau}]=(\frac{\partial r}{\partial \tau})^2[^4g_{rr}]+(\frac{\partial \theta}{\partial \tau})^2[^4g_{\theta\theta}]=0
\label{3-41}
\end{equation}

Eqs (\ref{3-40}) are satisfied if
\begin{equation}
M^2=a^2
\label{3-42}
\end{equation}
which means that the exterior solution is the extremal Kerr's solution,
while Eq (\ref{3-41}) is satisfied if as matching surfaces we chose the surfaces
\begin{equation}
be^{c\frac{x^2-1}{(r-M)^2}}-1=0
\label{3-43}
\end{equation}
The above equation gives two matching surfaces $S_\pm$ expressed by the relations
\begin{equation}
r_\pm^S=M \pm c' \sqrt{1-x^2}
\label{3-44}
\end{equation}
where
\begin{equation}
c'=\sqrt{\frac{c}{lnb}}>0
\label{3-45}
\end{equation}
Eqs (\ref{3-43}) and (\ref{3-44}) imply that the constants $b$ and $c$ must satisfy the relations
\begin{equation}
b>0 \>\>\>\mbox{and}\>\>\> \frac{c}{lnb}>0
\label{3-45'}
\end{equation}
Then combining Eqs (\ref{3-45'}) with the second of Eqs (\ref{3-28}) we find that we must have
\begin{equation}
c>0 \>\>\>\mbox{and}\>\>\> lnb>0
\label{3-45''}
\end{equation}
The matching surfaces $S_\pm$ of Eq (\ref{3-44}) are closed and axially symmetric. Since in  Boyer-Lindquist coordinates \cite{Bo2} the surface $r=constant$ is an oblate spheroid \cite{Gu1}, the matching surfaces of Eqs (\ref{3-44}) for $c' \ll M $ approximate oblate spheroids.

Since  relations $[K_{t\tau}]=[K_{\phi\tau}]=0$ are identically satisfied condition (\ref{3-39}) implies the relations \cite{Dr1}. \cite{Ky3}
\begin{equation}
[K_{tt}]=[g^{rr}]g_{tt,r}-[g^{\theta\theta}]R_{,\theta}g_{tt,\theta}=0
\label{3-46}
\end{equation}
\begin{equation}
[K_{t\phi}]=[g^{rr}]g_{t\phi,r}-[g^{\theta\theta}]R_{,\theta}g_{t\phi,\theta}=0
\label{3-47}
\end{equation}
\begin{equation}
[K_{\phi\phi}]=[g^{rr}]g_{\phi\phi,r}-[g^{\theta\theta}]R_{,\theta}g_{\phi\phi,\theta}=0
\label{3-48}
\end{equation}
\[[K_{\tau\tau}]=\frac{1}{2}(\frac{\partial r}{\partial \tau})^2\{[g^{rr}g_{rr,r}]+R_{,\theta}[g^{\theta\theta}g_{rr,\theta}]\}+\frac{\partial r}{\partial \tau}\frac{\partial \theta}{\partial \tau}\{[g^{rr}g_{rr,\theta}]-R_{,\theta}[g^{\theta\theta}g_{\theta\theta,r}]\}\]
\begin{equation}
-\frac{1}{2}(\frac{\partial \theta}{\partial \tau})^2\{[g^{rr}g_{\theta\theta,r}]+R_{,\theta}[g^{\theta\theta}g_{\theta\theta,\theta}]\}=0
\label{3-49}
\end{equation}
where all metric components refer to the 4-metric and we have used the notation $P_{,a}=\frac{\partial P}{\partial x^a}$. If Eqs (\ref{3-42}) and (\ref{3-43}), which imply the relations $[g^{rr}]=[g^{\theta\theta}]=0$, are satisfied we get
\begin{equation}
[K_{tt}]=[K_{t\phi}]=[K_{\phi\phi}]=0
\label{3-50}
\end{equation}
Also for $\tau=\theta$ we find that
\begin{equation}
[K_{\tau\tau}]=[K_{\theta\theta}]=\pm \frac{\sqrt{c lnb}}{(1-x^2)^{3/2}}
\label{3-51}
\end{equation}
where we have the plus sign if we choose $S_+$ as matching surface and the minus sign if we choose $S_-$ as such surface. Since for $S_+$ and $S_-$ we have $[K_{\theta\theta}]\neq 0 $ both surfaces are thin shells \cite{Ky4}.

 To calculate the surface energy tensor $S_i^j$ we use the Lanczos relation \cite{Is2}, \cite{Is3}, \cite{La1}
\begin{equation}
-8\pi S_i ^j=[K_{il}]^3g^{lj}-\delta_i^j([K_{ln}]^3g^{ln})
\label{3-52}
\end{equation}
For $^3g_{ij}$ and $K_{ij}$ given by Eqs (\ref{3-36}), (\ref{3-50}) and (\ref{3-51}) we find that the only non-vanishing $S_i^j$ are the following :
\begin{equation}
S_t ^t=S_\phi^\phi=\frac{1}{8\pi} [K_{\theta\theta}]^3g^{\theta\theta}= \frac{1}{8\pi}\frac{1}{^4g_{\theta\theta}+^4g_{rr}(\frac{\partial r}{\partial \theta})^2 }[K_{\theta\theta}]
\label{3-53}
\end{equation}
The  surface density $\sigma(\theta)$ is defined by the eigenvalue equation
\begin{equation}
 S_b ^au^b=-\sigma u^a \>\>\>\>\> \mbox{with} \>\>\>\>\>\> u_au^a=-1
\label{3-54}
\end{equation}
From  Eqs (\ref{3-44}), (\ref{3-51}), (\ref{3-53}) and (\ref{3-54}) we get
 \begin{equation}
\sigma(\theta)=-\frac{1}{8\pi}\frac{1}{^4g_{\theta\theta}+^4g_{rr}(\frac{\partial r}{\partial \theta})^2 }[K_{\theta\theta}]
\label{3-54'}
\end{equation}
which gives if we take as matching surface the surface $S_+$ of  Eq (\ref{3-44})
 \begin{equation}
  \sigma(\theta)=- \frac{\sqrt{c(lnb)^3}}{8\pi\sqrt{1-x^2}[(M\sqrt{lnb}+\sqrt{c(1-x^2)})^2+M^2 lnb x^2]}
\label{3-54''}
\end{equation}
and if we take the surface $S_-$ of  Eq (\ref{3-44})
 \begin{equation}
 \sigma(\theta) = \frac{\sqrt{c(lnb)^3}}{8\pi\sqrt{1-x^2}[(M\sqrt{lnb}-\sqrt{c(1-x^2)})^2+M^2 lnb x^2]}
\label{3-55}
\end{equation}
We see that if we take the surface $S_-$ as matching surface we get $\sigma(\theta)> 0$. Therefore for $M=a$ and certain values of its parameters the solution described in this section satisfies all energy conditions and matches to the extremal solution of Kerr on a surface which approximates an oblate spheroid and is a thin shell with positive surface density. The same thing happens for the family of solutions of Ref. \cite{Ky1} but for the case $M>a$ (non extremal case).

\section{Family of rotating anisotropic fluid solutions}

The non-zero components of the Ricci tensor of the family of rotating anisotropic fluid solutions, which has the metric of Eq (\ref{2-1}) with $f(r,x)$ and $h(r)$ given by Eqs (\ref{2-29}), are the following \cite{Bo3}
\begin{equation}
R_{rr}=\frac{2c(r-M)^2}{[(r-M)^2+a^2-M^2]^2}
\label{4-1}
\end{equation}
\begin{equation}
R_{r\theta}=\frac{2c(r-M)x
}{[(r-M)^2+a^2-M^2]sin\theta}
\label{4-2}
\end{equation}
\begin{equation}
R_{\theta\theta}=\frac{2cx^2}{1-x^2}
\label{4-3}
\end{equation}
The Ricci scalar of the solutions of the family is
\begin{equation}
R=\frac{2c[(r-M)^2+(a^2-M^2)x^2]}{b\rho^2}\{[(r-M)^2+a^2-M^2](1-x^2)\}^{-c-1}
\label{4-4}
\end{equation}
From the expressions (\ref{4-1}) - (\ref{4-3}) we calculate the matrix $R_\mu^\nu$ and then the eigenvalues $\lambda_i$,  $i=t, r, \theta, \phi$ of this matrix. We get
\begin{equation}
\lambda_t=\lambda_\theta=\lambda_\phi=0
\label{4-5}
\end{equation}
\begin{equation}
\lambda_r=\frac{2c[(r-M)^2+(a^2-M^2)x^2]}{b\rho^2}\{[(r-M)^2+a^2-M^2](1-x^2)\}^{-c-1}=R
\label{4-6}
\end{equation}
We see that Eq (\ref{1-1}) is satisfied, which means that the above solution is an anisotropic fluid solution.

From Eqs (\ref{4-5}) and (\ref{4-6}) we find that the eigenvalues $w_i$,  $i=t, r, \theta, \phi$ of the energy-momentum tensor $T_\mu^\nu=R_\mu^\nu-\frac{R}{2}\delta_\mu^\nu$ are given by the relations
\begin{equation}
w_t=w_\theta=w_\phi=-\frac{R}{2}
\label{4-7}
\end{equation}
\begin{equation}
w_r=\frac{R}{2}
\label{4-8}
\end{equation}
Therefore the energy density $\mu$ of the solutions is
\begin{equation}
\mu=-w_t=\frac{R}{2}
\label{4-9}
\end{equation}
and we have
\begin{equation}
\mu+w_r+w_\theta+w_\phi=0
\label{4-10}
\end{equation}
If $R>0$ we easily find that the solutions of the family satisfy the dominant the strong and therefore the weak energy conditions \cite{Ha2}.

The eigenvectors $(u_t)^\mu$, $(u_r)^\mu$, $(u_\theta)^\mu$ and $(u_\phi)^\mu$, which correspond to the eigenvalues $\lambda_t$, $\lambda_r$, $\lambda_\theta$ and $\lambda_\phi$ respectively are the following:
\begin{equation}
(u_t)^\mu=-\sqrt{\frac{\rho^2}{(r-M)^2+a^2x^2-M^2}}\delta_t^\mu
\label{4-11}
\end{equation}
\begin{equation}
(u_r)^\mu=\frac{1}{\sqrt{b\rho^2[(r-M)^2+x^2(a^2-M^2)]}}N^\frac{1-c}{2}\{(r-M)\delta_r^\mu+\frac{x}{\sqrt{1-x^2}}\delta_\theta^\mu\}
\label{4-12}
\end{equation}
\[(u_\theta)^\mu=\frac{1}{\sqrt{b\rho^2[(r-M)^2+x^2(a^2-M^2)]}}N^\frac{-c}{2}\{x[(r-M)^2+a^2-M^2]\delta_r^\mu-\]
\begin{equation}
(r-M)\sqrt{1-x^2}\delta_\theta^\mu\}
\label{4-13}
\end{equation}
\begin{equation}
(u_\phi)^\mu=\frac{1}{\sqrt{g_{\phi\phi}}}\delta_\phi^\mu
\label{4-14}
\end{equation}
where
\begin{equation}
N=(1-x^2)[(r-M)^2+a^2-M^2]
\label{4-15}
\end{equation}
From the above expressions we can calculate $(u_t)_\mu$, $(u_r)_\mu$, $(u_\theta)_\mu$ and $(u_\phi)_\mu$. We get
\begin{equation}
(u_t)_\mu=\sqrt{\frac{(r-M)^2+a^2x^2-M^2}{\rho^2}}\delta_{t\mu}
\label{4-16}
\end{equation}
\[(u_r)_\mu=\sqrt{\frac{b\rho^2}{[(r-M)^2+x^2(a^2-M^2)]}}N^\frac{1+c}{2}\{\frac{r-M}{(r-M)^2+a^2-M^2}\delta_{r\mu}+\]
\begin{equation}
\frac{x}{\sqrt{1-x^2}}\delta_{\theta\mu}\}
\label{4-17}
\end{equation}
\begin{equation}
[(u_\theta)_\mu=\sqrt{\frac{b\rho^2}{[(r-M)^2+x^2(a^2-M^2)]}}N^\frac{c}{2}\{x\delta_{r\mu}-
(r-M)\sqrt{1-x^2}\delta_{\theta\mu}\}
\label{4-18}
\end{equation}
\begin{equation}
(u_\phi)_\mu=\sqrt{g_{\phi\phi}}\delta_{\phi\mu}
\label{4-19}
\end{equation}
If we introduce again the notation of Eqs (\ref{3-29}) and use Eqs (\ref{2-1}), (\ref{2-29}),  (\ref{4-1}) - (\ref{4-4}), (\ref{4-7}) - (\ref{4-9}), (\ref{4-16}) and (\ref{4-17}) we find that the energy- momentum tensor $T_{\mu\nu}=R_{\mu\nu}-\frac{R}{2}g_{\mu\nu}$ of the family of solutions can be written in the form
\begin{equation}
T_{\mu\nu}=(\mu+w_\perp)(u_t)_\mu(u_t)_\nu+w_\perp g_{\mu\nu}+(w_\parallel-w_\perp)(u_r)_\mu(u_r)_\nu
\label{4-20}
\end{equation}
The above $T_{\mu\nu}$ has the form of the energy-momentum tensor of an anisotropic fluid \cite{He1} \cite{Le1}. From Eq (\ref{4-20}) we get the relation
\begin{equation}
R_{\mu\nu}=R(u_r)_\mu(u_r)_\nu
\label{4-21}
\end{equation}

If $M^2-a^2\geq 0$ the family of solutions has the infinite red shift surfaces
\begin{equation}
r_\pm ^{RS'} =M \pm \sqrt{M^2-a^2x^2}
\label{4-22}
\end{equation}
which are closed and axially symmetric.

The irremovable singularities of the solutions are the singularities of their invariants $R$ and $R^2$. The Ricci scalar $R$ of the family is given by Eq (\ref{4-4}), while the curvature scalar $R^2$ is given by the relation \cite{Bo3}
\begin{equation}
R^2=\frac{V}{b^2(\rho^2)^6}\{(r^2-2 M r+a^2)(1-x^2)\}^{-2-2c}
\label{4-23}
\end{equation}
where $V$ is a very complicated polynomial of $a$, $c$, $M$, $r$ and $x$. Therefore for all values of the parameter $c$ the invariants $R$ and $R^2$ are singular when $\rho^2=0$, which means that all solutions of the family have the ring singularity of Kerr's solution \cite{Ra1}, \cite{Vi1}. This is the only singularity of the solutions with $c\leq -1$, while the solutions with $c> -1$ are also singular if
\begin{equation}
r_\pm =M \pm \sqrt{M^2-a^2}
\label{4-24}
\end{equation}
and if
\begin{equation}
x=\pm 1
\label{4-25}
\end{equation}

To find if the family of solutions matches to the solution of Kerr on a surface $S'$ we shall examine if  Eqs (\ref{3-40}) and (\ref{3-41}), which come from the continuity of the first fundamental form, and  Eqs (\ref{3-46}) - (\ref{3-49}),
which come from the continuity of the second fundamental form, are satisfied. We easily find that  Eqs (\ref{3-40}) are satisfied. Also Eq (\ref{3-41}) is satisfied if
\begin{equation}
b[(1-x^2)(r^2-2 M r+a^2)]^c-1=0
\label{4-26}
\end{equation}
and if
\begin{equation}
(\frac{\partial r}{\partial \theta})^2\frac{1}{r^2-2 M r+a^2}+1=0
\label{4-27}
\end{equation}
Putting
\begin{equation}
\sqrt[c]{b}=\frac{1}{v}
\label{4-28}
\end{equation}
we get from Eq (\ref{4-26}) the matching surfaces $S'_\pm$
\begin{equation}
r_\pm ^{S'} =M \pm \sqrt{M^2-a^2+v(1-x^2)^{-1}}
\label{4-29}
\end{equation}
where the surface $S'_+$ ($S'_-$) is obtained from the above expression with the sign plus (minus).
Eqs (\ref{4-29}) imply that the constants must be chosen such that
\begin{equation}
M^2-a^2+v(1-x^2)^{-1}>0
\label{4-30}
\end{equation}
 Since we have for the surfaces $S'_+$ and $S'_-$  $[g^{rr}]=[g^{\theta\theta}]=0$ Eqs (\ref{3-46}) - (\ref{3-48}) are satisfied, while if we take $\tau=\theta$ we get
\begin{equation}
[K_{\tau\tau}]=[K_{\theta\theta}]=\pm \frac{c[M^2-a^2+v(1-x^2)^{-2}]^2}{[M^2-a^2+v(1-x^2)^{-1}]^{3/2}}
\label{4-31}
\end{equation}
where we have the plus sign if we choose $S'_+$ as matching surface and the minus sign if we choose $S'_-$ as such surface. Eq (\ref{4-31}) tells us that the surfaces $S'_+$ and $S'_-$ are thin shells.

Proceeding as in Sect III we find again that the non vanishing components of the surface energy tensor $S_i^j$ are given by  Eqs (\ref{3-53}) and the surface density $\sigma(\theta)$ by the first of  Eqs (\ref{3-55}). Then using  Eq (\ref{4-31}) we get for the surface $S'_+$
\begin{equation}
\sigma(\theta)= - \frac{c[M^2-a^2+v(1-x^2)^{-2}]}{8\pi[(r_+^{S'})^2+a^2x^2][M^2-a^2+v(1-x^2)^{-1}]^{1/2}}
\label{4-32}
\end{equation}
and for the surface $S'_-$
\begin{equation}
\sigma(\theta)= \frac{c[M^2-a^2+v(1-x^2)^{-2}]}{8\pi[(r_-^{S'})^2+a^2x^2][M^2-a^2+v(1-x^2)^{-1}]^{1/2}}
\label{4-32'}
\end{equation}
where $r_\pm ^{S'}$ is given  Eq (\ref{4-29}). Therefore since relation (\ref{4-30}) must hold it is  $\sigma(\theta)>$ for the surface $S'_+$ if $c[M^2-a^2+v(1-x^2)^{-2}]<0$ and for the surface $S'_-$ if $c[M^2-a^2+v(1-x^2)^{-2}]>0$.

 Eq (\ref{4-27}) has a solution only if $M^2 > a^2$ and this solution is
\begin{equation}
r_\pm ^{S''}=M  \pm\sqrt{M^2-a^2}cos(\theta-\theta_0)
\label{4-35}
\end{equation}
where $\theta_0$ is an arbitrary constant. The matching surfaces ${S''}_\pm$ of this solution are closed, axially symmetric and as in the case of the anisotropic fluid solution of Sect III if $\sqrt{M^2-a^2} \ll M $ approximate oblate spheroids.

From  Eqs (\ref{4-22}) and (\ref{4-35}) for $\theta_0=0$ we get
\begin{equation}
r_+^{RS'}\geq r_+^{S''}
\label{4-36}
\end{equation}
where the equality holds for $x=1$. Also since the event horizons of Kerr's solution are given by the relations
$r_\pm^H=M\pm\sqrt{M^2-a^2}$ we get
\begin{equation}
{r}_+^H \geq r_+^{S''}
\label{4-37}
\end{equation}
Proceeding as before we can calculate $ [K_{ij} ]$, $S_i^j$ and $\sigma(\theta)$ for the matching surfaces ${S''}_\pm$. However the calculations involved are very long and for this reason we shall not give $\sigma(\theta)$ explicitly. Also since in this case the condition (\ref{3-39}) is not satisfied for all $i$ and $j$ the surfaces ${S''}_\pm$ are thin shells.


\begin{thebibliography}{99}
\bibitem{Ky3} Since the right hand side of Eqs (\ref{3-46})-(\ref{3-49}) is zero a positive normalization factor of the unit vector, with the covariat derivative of which we define the extrinsic curvature, is ommitted.  We shall ommit this factor everywhere in the paper.
\end{thebibliography}
\end{document}